# Theory of the size effect of the properties of the relaxor ferroelectric films.

E.A.Eliseev*, M.D.Glinchuk

*Institute for Problems of Materials Science, National Academy of Science of Ukraine,*

*Krjijanovskogo 3, 03142 Kiev, Ukraine,*

*eliseev@mail.i.com.ua

For the first time we proposed the model for the calculations of the relaxor ferroelectrics films properties in the framework of the random field theory. We took into account the misfit strain between film and substrate as well as surface piezoelectric effect that causes built-in electric field in the strained films. In the statistical theory framework we calculated random field distribution function with the electric dipoles and monopoles as the field sources. It was shown that with thickness decrease the mean field decreases, while the width of the distribution function increases. This leads to the additional smearing of the phase transition in the films in comparison to the bulk relaxors. As an example the dependence of the order parameter and dielectric susceptibility on the film thickness, temperature and random fields distribution function parameters was obtained. For free standing film the existence of critical thickness of relaxor state transformation into glassy state was predicted. Contrary to this the appearance of misfit strain induced ferroelectric phase appeared to be possible for some pairs film-substrate. We have shown that susceptibility temperature maximum shift with frequency in relaxor ferroelectric thin films obeys Vogel-Fulcher law with parameters dependent on film thickness. For the first time the analytical dependences of freezing temperature decreases and activation energy on the thickness was obtained, namely freezing temperature decreases and activation energy increases with film thickness decrease. Obtained results quantitatively agree with the available experimental data for $PbMg_{1/3}Nb_{2/3}O_3$ relaxor thin films.

Keywords: relaxor ferroelectric films, random electric field, misfit strain.

PACS: 77.80.-e, 77.55.+f, 77.22.Gm

## 1. INTRODUCTION

Relaxor ferroelectric materials have the prominent dielectric, electromechanical and electrooptical properties that makes them attractive for both the numerous applications and fundamental study. In particular relaxor ferroelectrics $PbMg_{1/3}Nb_{2/3}O_3$ (PMN), $PbZn_{1/3}Nb_{2/3}O_3$ (PZN) and their solid solutions with normal ferroelectrics $PbTiO_3$ (PT) have been the focus of intensive studies in view of their applications in actuators and transducers. Because the single crystals of relaxor ferroelectrics revealed superior properties [1], [2] much better than bulk ceramics thin films as the more cheaper and being able to integrate into the modern semiconductors technology were considered as promising candidates for various microelectromechanical applications. However up to now these expectations did not appeared true. Moreover the properties of the films appeared to be dependent on the technology of the deposition process, type of substrate and electrodes. In many cases different methods of the analysis of experimental data lead to the contradictions between the results for the same relaxor ferroelectrics obtained by different authors (see e.g. [3], [4], [5], [6], [7], [8], [9], [10]). The attempts to understand the reason of these contradictions and to find the ways and conditions to advance the properties of relaxor films to the ones of single crystals are continuing up to now (see e.g. [11]). The main still opened question is what properties could be expected in thin films of relaxor ferroelectrics? The absence of the generally recognized theoretical models of relaxor ferroelectric films able to describe the observed experimental data does not allow to give answer to this question.

The description of the normal ferroelectrics thin film properties including size driven phase transitions or its absence in the ultrathin films was successfully performed recently allowing for depolarization field and mismatch strain between film and substrate [12], [13], [14]. In these and many other papers (see e.g. [15]) the films were considered in the phenomenological theory approach on the basis of the free energy functional expansion. The strong influence of the random field, inhomogeneity and non-ergodicity of the system makes it cumbersome to use free energy even for the description of the bulk relaxor ferroelectrics (see e.g. [16]). It is obvious that some special model for relaxor ferroelectric films should be developed.

In this paper we proposed for the first time the random field based model of relaxor ferroelectric films properties size effect calculations. In this model developed similarly to the model of bulk relaxors [17] we consider the system of dipoles randomly distributed in the host lattice (paraelectric Burns reference phase) along with other sources of random field. Embedded dipoles tend to order system with temperature decrease so that in the absence of random electric field the ferroelectric long-range order could appear below Burns temperature $T_d$. Because of the random field the system transforms into relaxor ferroelectrics. In order to obtain the properties of the relaxor ferroelectric in the framework of our model one should find the properties of the Burns phase depending on electric field. Then the properties of relaxor ferroelectrics can be found by averaging over the random electric field with the random field distribution function. Some preliminary results had shown [18] that this function for the thin films has several peculiarities related to the influence of the surfaces.

Since the misfit between film and its substrate is one of the essential factors which affects the film properties we took into account the misfit strain arising due to the mismatch between lattice constants, thermal expansion coefficients of film and its substrate. It was shown [14], [19], [20] for the ordered ferroelectrics that due to the broken symmetry on the surface of the ferroelectric films this strain leads to the appearance of built-in internal field. In this paper we applied these results to the description of relaxor ferroelectric films.

The comparison of the obtained results with available experimental data has shown that the theory fits rather well the observed properties of relaxor ferroelectric films.

## 2. THE MODEL

Hereinafter we consider epitaxial film with thickness $h$ on the substrate with top and bottom electrodes (see Fig. 1)

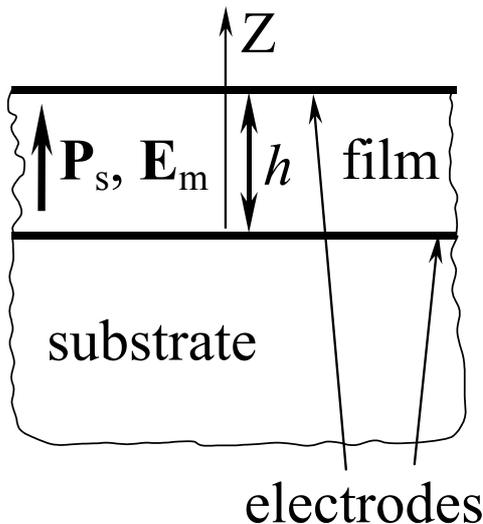

**Figure 1**. Film of relaxor ferroelectrics is placed between two electrodes and lay on the thick substrate. Polarization, electric field and Z-axis are pointed perpendicularly to the film surfaces.

In the heterostructure like considered one mismatch between in-plain lattice constants and thermal expansion coefficients of the bulk of the film, electrodes and substrate causes misfit strain in the considered system. Since for the not very high deposition temperatures the main contribution to the misfit strain is related to the lattice constants difference hereinafter we neglected thermal expansion. The substrate usually is much thicker than the film and electrodes (see Fig. 1) so it remains unstrained and film and electrodes became homogeneously stretched or compressed so that they have the same in-plain lattice constants as that of the substrate (see e.g. [14], [19], [20], [21]). In this case the misfit strain can be found as

$$U_m = \frac{b-a}{b}, \quad (1)$$

where $b$ and $a$ is the in-plain lattice constants of the substrate and freestanding film respectively.

It should be noted that if the film thickness exceeds some critical value $h_{md}$, then misfit dislocation appears in film in order to compensate the misfit strain. It was shown [22], that this effect could be taken into account by using the effective substrate lattice constant $b^*$ when calculating the effective misfit strain $U_m^* = (b^* - a)/b^*$. Recently [20] we proposed the approximate expression for the effective strain:

$$U_m^*(h) = \frac{b^*(h) - a}{b^*(h)}, \quad b^*(h) = \begin{cases} b\left(1 - U_m\left(1 - \frac{h_{md}}{h}\right)\right), & h > h_{md} \\ b, & h \leq h_{md} \end{cases} \quad (2)$$

Here the critical thickness of the misfit dislocations generation $h_{md}$ decreases with misfit strain $U_m$ increase [22].

When considering the electric field of the different type sources inside the film one can neglect the substrate influence, since electrodes are usually much thicker than Thomas-Fermi screening length (which is of lattice constant order) and effectively shield the electric field between them.

In framework of relaxor ferroelectric model based on the random field theory [17] one should find the distribution function of the random electric field and the physical properties dependence on the electric field for the system of electric dipoles in host lattice when all the sources of random electric field are absent. For the films all the physical properties are inhomogeneous and depend on the distance from the surfaces, i.e. the coordinate $z$ (see e.g. [23], [24]). Then observable physical quantity $\langle A(z,T) \rangle$ of disordered (relaxor) ferroelectrics can be found with the help of statistical averaging of the physical quantity of the reference phase on the random electric field:

$$\langle A(z,T) \rangle = \int_{-\infty}^{+\infty} f(\vec{E},z) A(\vec{E},z,T) d\vec{E} \quad (3)$$

Here $f(\vec{E},z)$ is the distribution function of the random electric field $\vec{E}$, $A(\vec{E},z,T)$ is the observable physical quantity of the system of dipoles without the random field. The dependence of $f(\vec{E},z)$ on coordinate is related to the influence of the film surfaces.

In the next section we proceed to calculation of the random electric field distribution function for the films of relaxor ferroelectrics.

## 3. RANDOM FIELD DISTRIBUTION FUNCTION

The distribution function of the random electric field created by the different type sources was calculated in the statistical theory framework earlier for bulk relaxor ferroelectric [25]. Some preliminary results related to the random field distribution function for the films of relaxor ferroelectrics can be found in [18], [26]. In this case the contribution of the surfaces was taken into account by the method of the image charges [27]. The electric field for the each type of sources was calculated for the film between two metallic electrodes.

The distribution function of the random field can be introduced by the following way [25]:

$$f(\vec{E}) = \overline{\langle\langle \delta(\vec{E} - \vec{E}(\vec{r}_i))\rangle\rangle} \qquad (4)$$

where $\vec{E}(\vec{r}_i)$ is local electric random field in the point $\vec{r}_i$, which in general case includes the contributions of both the real random field sources and its images. The bar denotes averaging over spatial configurations of random field sources, $\langle\langle \ldots \rangle\rangle$ means thermal averaging over orientations of random dipoles so that the distribution function is expressed through itself in a self-consistent manner. The calculation of $f(\vec{E})$ on the basis of expression (4) was carried out with the help of the statistical method [25]. It yields

$$f(\vec{E}) = \frac{1}{(2\pi)^3} \iiint \exp\left(i\vec{\rho}\cdot\vec{E} - \sum_k F_k(\vec{\rho})\right) d^3\rho \qquad (5)$$

$$F_k(\vec{\rho}) = n_k \int_V \langle\langle \exp(-i\vec{\rho}\cdot\vec{E}_k(\vec{r})) - 1 \rangle\rangle d^3r \qquad (6)$$

where $\vec{E}_k(\vec{r})$ is an electric field in the point $\vec{r}$, created by $k$-th type of random field sources with concentration $n_k$, all the sources are supposed to be independent.

Hereinafter we consider two types of sources, namely randomly distributed monopoles and dipoles with concentrations $n_m$ and $n_d$ respectively. Also we suggest that dipoles can be directed either along or against *z*-axis. That is why we can choose vector $\vec{\rho}$ as follows $(0,0,\rho)$ and consider the random electric field oriented along z-axis because the field components perpendicular to dipoles cannot influence their behavior.

It is easy to show that for the high enough concentration of random field sources the Gaussian approximation can be used for the calculation of the distribution function characteristics. Namely, the real and imaginary parts of expression (6) can be represented in the form:

$$\mathrm{Im}(F_k(\rho)) \approx \rho\, n_k \langle\langle \int_V E_{kz}(\vec{r}) dV \rangle\rangle \equiv E_{0k}\, \rho \qquad (7)$$

$$\mathrm{Re}(F_k(\rho)) \approx \rho^2 \frac{n_k}{2} \langle\langle \int_V E_{kz}(\vec{r})^2 dV \rangle\rangle \equiv \Delta_k\, \rho^2 \qquad (8)$$

The thermal averaging $\langle\langle \ldots \rangle\rangle$ over orientations in Eq. (8) is trivial for the dipoles with two possible orientations since the square of the field is the same for the two different orientations.

The dependence of contribution (8) of the different type sources to the distribution function width on the coordinate inside the film of relaxor ferroelectrics was obtained recently [26]. For the arbitrary distance from the film surfaces expressions (8) was calculated numerically. Also we proposed an approximate approach for the relaxor ferroelectrics films with the large enough thickness. We took into account the source images from the bottom and upper surface separately. Allowing for the electric field decrease with distance we do not take into account the other images and developed the following approximate formula:

$$\Delta_k(z,h) \cong \frac{\Delta_{kS}(h/2-z)\Delta_{kS}(h/2+z)}{\Delta_k^\infty}. \qquad (9)$$

Here function $\Delta_{kS}(a)$ determines the contribution to the width for of the source of type *k* at the distance *a* from the flat boundary between dielectric and metallic electrode, $\Delta_k^\infty$ is the bulk value of $\Delta_{kS}(a)$.

It is found that the exact expression for $\Delta_k$ which takes into account the eight nearest images of the source as well as their interactions can be approximated by Eq. (9) within the accuracy of several percents for the films with thickness higher than several lattice constants.

For the monopoles with charge *q* that is distributed inside the sphere with radius *b* we obtained [18]:

$$\Delta_{mS}(a) = \Delta_m^\infty \left[1 - \frac{b^3}{8a^3}\right], \quad \Delta_m^\infty = \frac{2\pi}{3} \frac{n_m}{b} \left(\frac{q}{\varepsilon}\right)^2. \tag{10}$$

where $\varepsilon$ is the dielectric permittivity of the host lattice.

For the dipoles with the arm much smaller than lattice constant one should take into account indirect interaction between embedded dipoles via soft mode phonons of host lattice [25]. In this case even the calculation of $\Delta_{dS}(a)$ was performed numerically. At the large distance $a$ from the surface it can be easily expressed in terms of elementary functions:

$$\Delta_{dS}(a) = \Delta_d^\infty \left(1 + \frac{15}{8}\left(\frac{r_c}{a}\right)^3 \left[1 - 3\left(\frac{r_c}{a}\right)^2\right]\right), \quad \Delta_d^\infty = \frac{8\pi}{15} \frac{n_d}{r_c^3}\left(\frac{d^*}{\varepsilon}\right)^2 \tag{11}$$

Here $d^*$ is the effective dipole moment, $r_c$ is the correlation radius of the host lattice polar phonons fluctuations.

Using Eqs. (9)-(11) the width of the random electric field distribution function $\Delta E$ can be written as:

$$\Delta E(z) = \sqrt{\Delta_m(z) + \Delta_d(z)}. \tag{12}$$

Next we consider the contributions (7) of different type sources to the mean value of the random electric field. It was shown that the monopoles contribute nothing to the mean field ($E_{0m} = 0$) in the films [26] as well as in the bulk systems [25]. The contribution of dipoles $E_{0d}$ is proportional to the fraction of coherently oriented dipoles $L$:

$$E_{0d} = L E_0 \tag{13}$$

For the bulk system $E_0$ has the form [25]:

$$E_0^\infty = \frac{4\pi n_d d^*}{\varepsilon} \tag{14}$$

For the films $E_0$ will be dependent on the film thickness, misfit strain, correlation energy and the depolarization field strength. Since we consider the dipoles with two possible orientations this system without random field represents ferroelectrics of order-disorder type. The films of this type material were considered earlier [23] for the dipoles perpendicular to the film surfaces allowing for the depolarization field and the correlation energy.

However in this model effects of mismatch between film and substrate due to the different lattice constant was not taken into account since thick enough films can be considered on the base of the bulk free energy with coefficients dependent on the misfit strain, elastic and electrostriction constants [21]. For the thin films along with mismatch effects one should take into account size effects, depolarization field influence (see e.g. [14], [19]) and misfit strain relaxation due to the dislocation generations [20].

It can be shown, that the thermal averaging in (7) with Hamiltonian used in [23], [24] for the description of films of order-disorder ferroelectrics will lead to the following expression for the mean field of the relaxor ferroelectric film:

$$E_0(h) = E_0^\infty \left(1 - \frac{2\delta}{2l_d^2 + h(\lambda + l_d)}\right) + n_d d^* \frac{2Q_{12} U_m^*(h)}{s_{11} + s_{12}}, \quad l_d \approx \sqrt{\delta/\varepsilon} \qquad (15)$$

Here $\delta$ determines the correlation energy, $\lambda$ is the extrapolation length, $Q_{12}$ is the component of electrostriction tensor, $U_m^*(h)$ is the effective strain from Eq.(2), $s_{11}, s_{12}$ is the elastic constants. The first term is related to the surface and correlation effects allowing for the depolarization field [23], the second is related to the electrostriction coupling between the polarization and strain [14], [19], [20], [21].

It is seen from Eq. (15) that with thickness increase $E_0(h)$ tends to the bulk value $E_0^\infty$. In the framework of our model quantity $d^* E_0(h)/k_B$ equals to so called Burns temperature $T_d$ [17], [28] at which the transition from paraelectric to the ferroelectric phase could exist in the absence of the random electric field. It is found experimentally that the polar nano regions arise in the relaxor ferroelectrics below the temperature $T_d$ (see [28] and ref. therein). Since the mean field (15) for the films depends on thickness our theory predicts the size effect for the Burns temperature of relaxor films similar to the size effect of normal ferroelectrics.

Using the definition (5), width (9)-(12) and the mean field (13), (15), the distribution function $f(E, z)$ can be written as follows:

$$f(E, z) = \frac{1}{2\sqrt{\pi}\,\Delta E(z)} \exp\left(-\frac{(E - E_0(h)L)^2}{4\Delta E(z)^2}\right) \qquad (16)$$

This function depends on the order parameter of the system of dipoles $L$, which should be found by the averaging (3) with distribution function (16) the fraction of coherently oriented dipole of the reference phase.

The obtained distribution function (16) permits to calculate all the properties of relaxor ferroelectric film that we demonstrate in the next section.

# 4. RELAXOR FERROELECTRIC FILM PROPERTIES.

## 4.1 Order parameter and susceptibility of relaxor ferroelectric film

The distribution of fraction of coherently oriented dipoles $l(z,E)$ inside the film of reference phase can be found as follows [23], [24]:

$$l(z,E) = (1-\varphi(z))\tanh\left(\frac{d^*}{k_B T}(E + E_m(h))\right),$$
$$\varphi(z) = \frac{\cosh(z/l_d)}{\cosh(h/2l_d) + \sinh(h/2l_d)\lambda/l_d}, \quad z \in \left[-\frac{h}{2}, \frac{h}{2}\right]. \quad (17)$$

Here $E$ is the electric field, $E_m(h)$ is the built-in internal misfit induced field. It was shown [14], [19], [20] that due to the broken symmetry on the film surface the surface piezoelectric effect arises and so misfit strain between film and its substrate induces field $E_m(h)$. This field depends on the film thickness and decreases with thickness increase, namely:

$$E_m(h) = E_U \frac{\delta}{2l_d^2 + h(\lambda + l_d)}, \quad E_U = \frac{4\pi}{\varepsilon}\frac{U_m}{d_{31}}. \quad (18)$$

Here $d_{31}$ is component of the tensor of surface piezoelectric effect [14], [19], [20], $U_m$ is the misfit strain from Eq.(1). Internal field (18) smears the phase transition, shifts hysteresis loop and causes the self-polarization in the thin ferroelectric films [14], [19].

For the films of relaxor ferroelectrics one has to average quantity (17) on the random electric field created by different sources [25]. Taking into the distribution function (16) and expressions (3), (15)-(18), we can write the following equation for the order parameter $L$ spatial distribution:

$$L(T,z) = \int_{-\infty}^{+\infty} \frac{1-\varphi(z)}{2\sqrt{\pi}\Delta E(z)} \exp\left(-\frac{E^2}{4\Delta E(z)^2}\right) \tanh\left(\frac{d^*}{k_B T}(E + E_{ext} + E_m(h) + E_0(h)L(T,z))\right) dE. \quad (19)$$

Here $E_{ext}$ is the external electric field pointed along z-axis.

One can see from Eq. (19) that in the films additional inhomogeneity of order parameter related to the influence of surface on the distribution function appears and conserves after averaging over random field that was the only source of the properties inhomogeneity in the bulk relaxors. It is obvious that all the physical properties related to the order parameter should have the same inhomogeneity in the films. In particular with the help of expression (19) we can calculate linear dielectric susceptibility $\chi = n_d d^* \left(\partial L/\partial E_{ext}\right)_{Eext=0}$.

$$\chi(T,z) = n_d d^* \frac{\partial L(T,z)}{\partial E_{ext}}\bigg|_{Eext=0} = \frac{n_d (d^*)^2 I(T,z)}{k_B T - d^* E_0(h) I(T,z)};$$

$$I(T,z) = \int_{-\infty}^{+\infty} \frac{1-\varphi(z)}{2\sqrt{\pi}\,\Delta E(z)} \exp\left(-\frac{E^2}{4\Delta E(z)^2}\right) \mathrm{sech}\left(\frac{d^*}{k_B T}(E + E_m(h) + E_0(h) L(T,z))\right)^2 dE$$

(20)

Using the distribution function width dependence on the coordinate inside the film, one can obtain with the help of Eqs. (18), (19) the distribution of order parameter and susceptibility inside the film. Then after averaging over coordinate z we calculated the observable quantities

$$\overline{A}(T) = \frac{1}{h}\int_0^h A(T,z)dz \qquad (21)$$

namely spontaneous order parameter $\overline{L}$ (at $E_{ext}=0$) and static susceptibility $\overline{\chi}$. Their dependence on temperature is represented in Figs. 2-4 for the different values of the film thickness, mismatch induced internal field and the distribution function width $\Delta E$ that determines the degree of disorder of relaxor ferroelectrics [17], [28].

It is seen from Fig. 2 that for the ideal case of the freestanding film ($E_m(h)=0$) there is critical values of the temperature and film thickness at which order parameter vanishes, films transmits from the mixed ferroglass phase to the dipole glass state [28] and susceptibility has sharp maximum.

For the fixed temperature, film thickness and misfit strain $\overline{L}$ value decreases with degree of disorder increase; the largest $\overline{L}$ value being characteristic for the completely ordered film (curves 1).

The critical temperature decreases with width $\Delta E$ increase or with thickness decrease so that for the thinnest film and large distribution width (see curves 5 in Figs. 2a, 2c) the spontaneous order parameter is equals to zero and susceptibility slowly decreases with temperature increase.

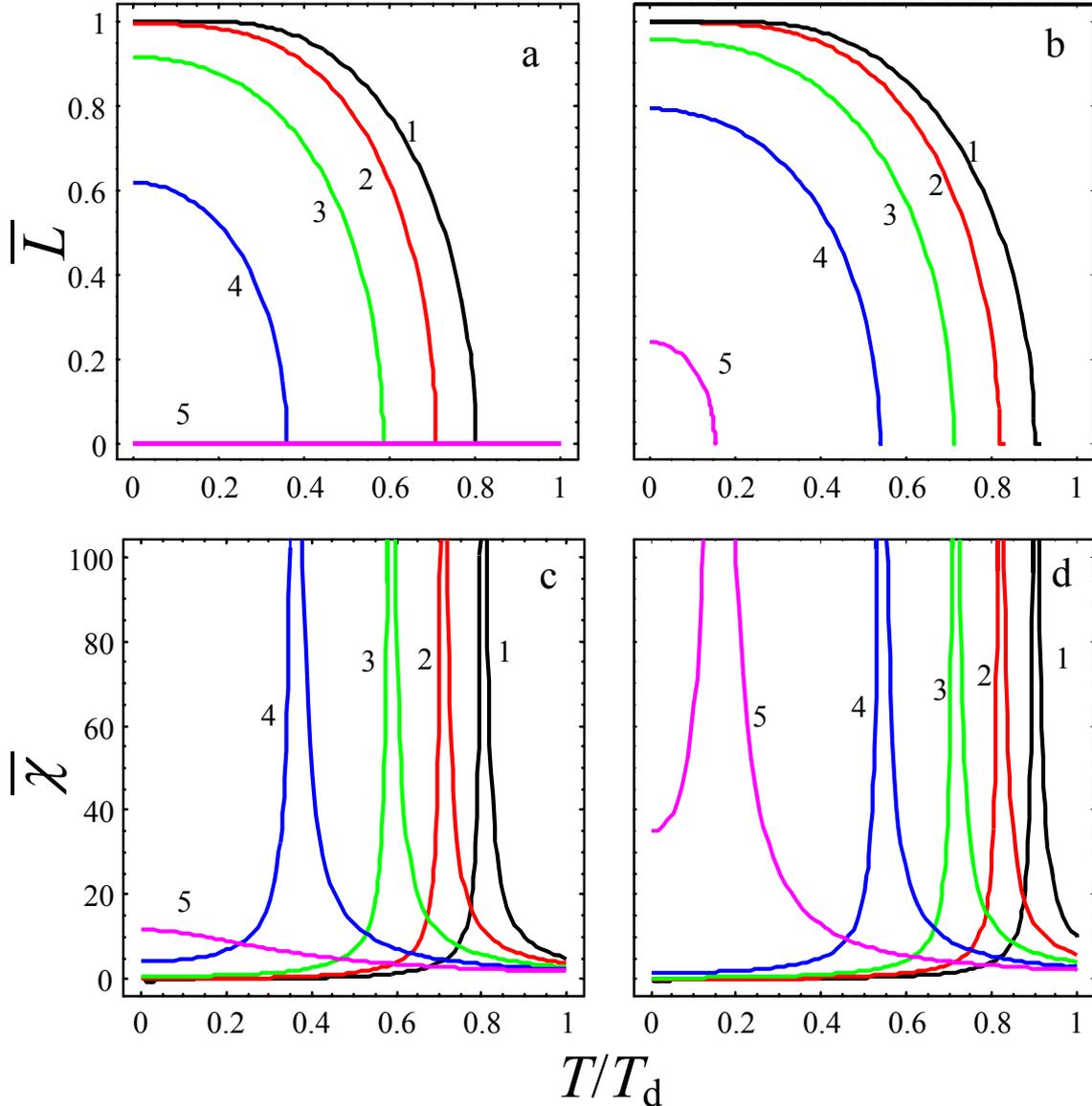

**Figure 2**. The order parameter (a, b) and static susceptibility (c, d) dependence on temperature for the free-standing film. The following parameters are used $\Delta E/E_0 = 0, 0.4, 0.6, 0.8, 1$ (curves 1, 2, 3, 4, 5) $h/l_d$=100 (a, c), 200 (b, d), $E_U/E_0^\infty = 0$.

For the strained films the transition smears, susceptibility maximum diffuses $\overline{\chi}$ and order parameter $\overline{L}$ do not vanish at the critical point due to the internal field (18) influence (see Figs. 3 and 4). It is seen that the increase of this field amplitude leads to the increase of order parameter and to the decrease of susceptibility (compare Fig. 3 and Fig. 4). The field $E_m(h)$ induces non-zero polarization $\overline{L} \neq 0$, i.e. the ferroglass phase appearance for the thinner films with high degree of disorder ($\Delta E \sim E_0$, see curve 5 in Fig. 4a), which would be in the dipole glass state under the conditions of free standing film (see curve 5 in Fig. 2a); at the same time the susceptibility for these film have no maximum (see curves 5 in Figs. 2c).

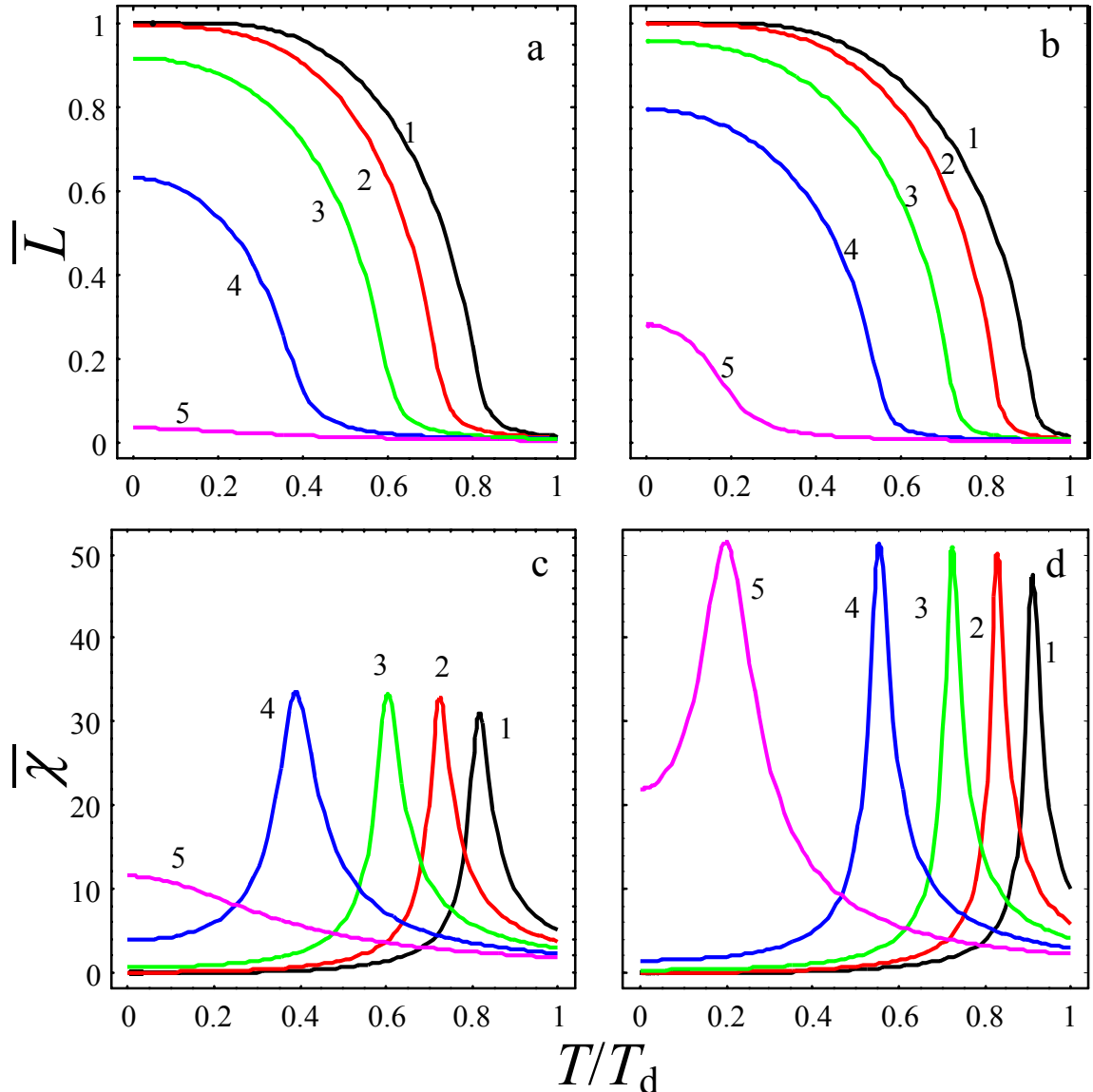

**Figure 3**. The order parameter (a, b) and static susceptibility (c, d) dependence on temperature for the slightly strained film. The following parameters are used $\Delta E/E_0 = 0, 0.4, 0.6, 0.8, 1$ (curves 1, 2, 3, 4, 5) $h/l_d$=100 (a, c), 200 (b, d), $E_U/E_0^\infty = 0.03$.

Qualitatively the same behavior one can see for the relaxor ferroelectrics with intermediate degree of disorder (compare the curves 2, 3 in the Figs. 3, 4). Since the curves 2 and 3 tend to the curve 1 for the ordered ferroelectric film with $E_U/E_0^\infty$ increase it is not excluded the transformation of relaxor ferroelectric film into completely ordered ferroelectric film due to the influence of mismatch induced electric field. Note that such type of transformation was observed in bulk relaxor ferroelectrics under the influence of external electric field [29]. For the special choice of relaxor film –substrate pair such phenomenon can be expected without external filed application.

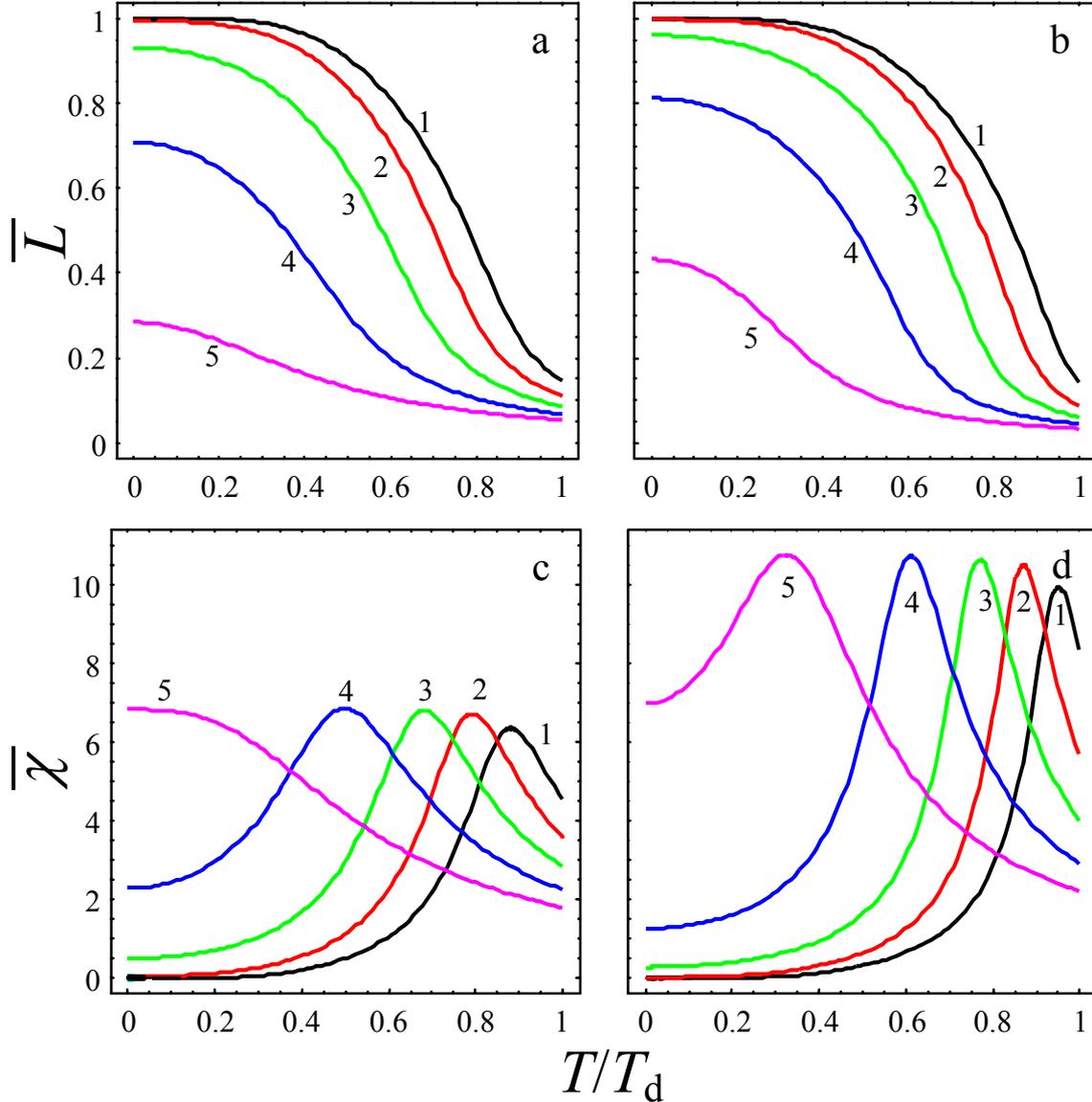

**Figure 4**. The order parameter (a, b) and static susceptibility (c, d) dependence on temperature for the strained film. The following parameters are used $\Delta E/E_0 = 0, 0.4, 0.6, 0.8, 1$ (curves 1, 2, 3, 4, 5) $h/l_d$=100 (a, c), 200 (b, d), $E_U/E_0^\infty = 0.3$.

The temperature $T_f$ corresponding maximum of dielectric susceptibility can be found from the following equation:

$$\frac{\partial \bar{\chi}(T)}{\partial T} = 0 \quad \Rightarrow \quad \left(T\frac{\partial \bar{I}(T)}{\partial T} - \bar{I}(T)\right)\bigg|_{T=T_f} = 0. \quad (22)$$

Because of the complex form of integral $\bar{I}(T)$ (see Eq. (20) and (21)) Eq. (22) can be solved only numerically.

## 4.2 Modified Vogel-Fulcher law

Mismatch induced field influences not only the static polar and dielectric properties of strained ferroelectric films, but also the dynamic dielectric response. Despite of the fact that the calculation of the dielectric susceptibility for the alternating external field was out of the scope of our paper we can analyze some general features of the dynamic response without the detailed calculations of dielectric spectrum. One of the main peculiarities of the dynamic dielectric properties of relaxor ferroelectrics is the pronounced dependence of the temperature of maximum of dielectric permittivity $T_m$ on the external field frequency $\omega$ (see, e.g. [28]). Usually it obeys so called Vogel-Fulcher (V-F) law that was obtained in [30] for the bulk relaxor ferroelectrics in the framework of random field theory allowing for barriers of dipoles reorientations dependence on the random field. We suggest that for the thin films one should take into account the influence of the internal built-in field on the height of the barrier between possible dipoles orientations. Since for the conventional V-F law at $\omega \to 0$ temperature of susceptibility maximum tends to freezing temperature one can expect that the freezing temperature coincides with the temperature $T_f$ at which the static susceptibility $\bar{\chi}$ from Eq. (20) has maximum. Thus the activation energy and the freezing temperature will be thickness dependent and V-F law for the film with thickness $h$ will have the form:

$$\omega = \omega_0 \exp\left(-\frac{T_a(h)}{T_m - T_f(h)}\right), \quad T_a(h) = \frac{E_a + d^* E_m(h)}{k_B} \tag{23}$$

Here $T_a(h)$ is the activation energy in temperature units, $E_a$ is the activation energy of the bulk system, $T_f(h)$ is the freezing temperature determined by Eq. (22).

The dispersion law (23) conserves the conventional form of V-F law for bulk relaxors but with the renormalized parameters dependent on the film thickness. The possibility of the description of the ferroelectric film properties with the help of the conventional free energy with renormalized expansion coefficients dependent on the film thickness was shown recently [24], [12], [13].

## 5. COMPARISON WITH EXPERIMENT

Now we will compare the obtained results with experimental data for the films of relaxor ferroelectrics. Unfortunately in the most of the papers there were no measurements of the size dependence of the physical properties because the measurements were usually performed on the one film or several films with the same thickness (see e.g. [3], [4], [5], [7], [8], [9], [10]).

From this point of view the most interesting results were published in Ref. [6]. Author of Ref. [6] studied the structure, dielectric, polar and electromechanical properties of the lead magnesium niobate $PbMg_{1/3}Nb_{2/3}O_3$ (PMN) films with different thickness on the substrates of $TiO_2/Pt/TiO_2/SiO_2/Si$ and $PbTiO_3/Pt/TiO_2/SiO_2/Si$ with $TiO_2$ and $PbTiO_3$ as seeding layers, Pt as bottom electrode. They also compared obtained results with properties of bulk ceramics of PMN. The dielectric response for the different external field frequencies was investigated in the broad temperature range. It was found, that the permittivity is several times smaller than that of ceramics and single crystals, its maximum is shifted to the higher temperatures. Piezoelectric properties and large pyroelectric effect were revealed in the as prepared films without poling treatment as well as asymmetry in the hysteresis loop. These facts speak in favor of the presence of large internal field, which polarize the considered films, stiffens dielectric response and shifts hysteresis loop. Therefore we can apply our model to the description of the some experimental data from Ref. [6].

We fit the temperatures of permittivity maxima to the V-F law (23) in order to obtain the freezing temperature and activation energy values did not reported by authors [6] for the three films with different thickness (see boxes in Fig. 5).

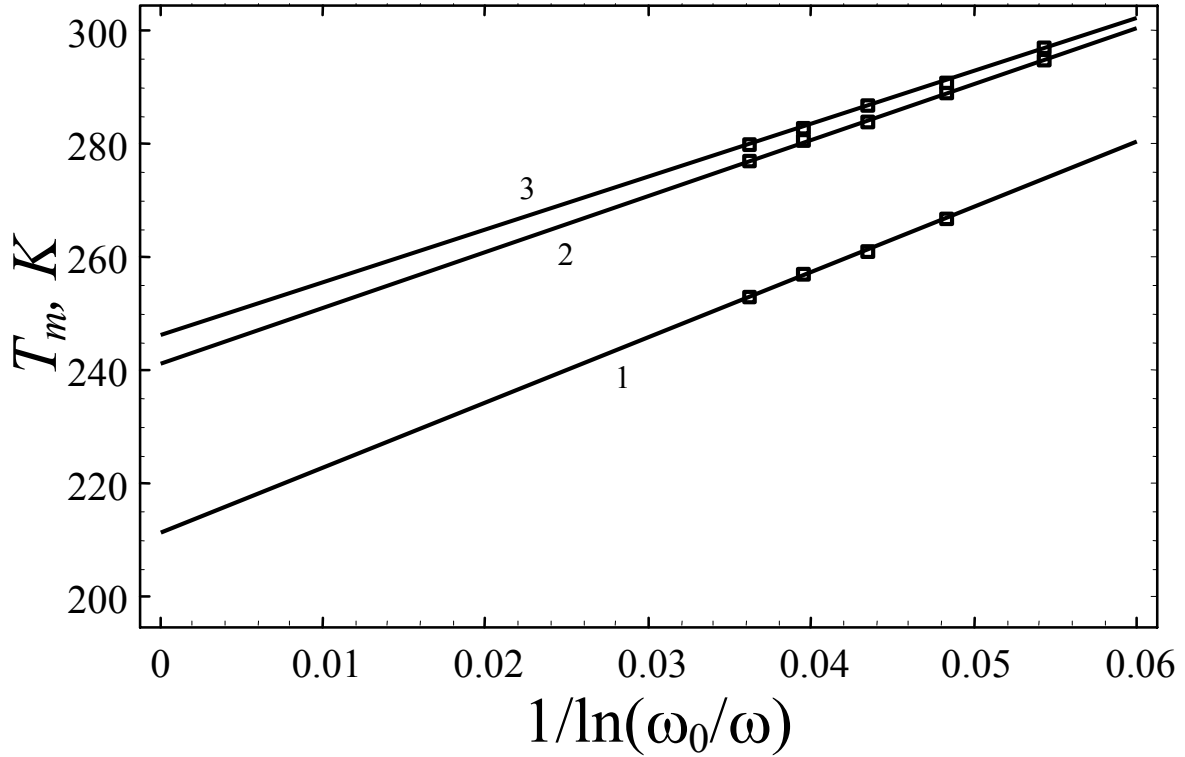

**Figure 5**. Vogel-Fulcher fit (solid lines) to the experimental data [6] of the temperature of dielectric response maxima dependence on the external electric field frequency (boxes) obtained for the $PbMg_{1/3}Nb_{2/3}O_3$ films of different thickness 430, 510, 770 nm (curves 1, 2, 3 respectively), $\omega_0 = 10^{14}\ Hz$.

Then we compare the obtained values for $T_a(h)$ and $T_f(h)$ with the proposed expression (23) with respect to Eqs. (18) and (22). It is seen from Fig. 6 that the obtained theoretical dependence fits the experimentally observed values rather well.

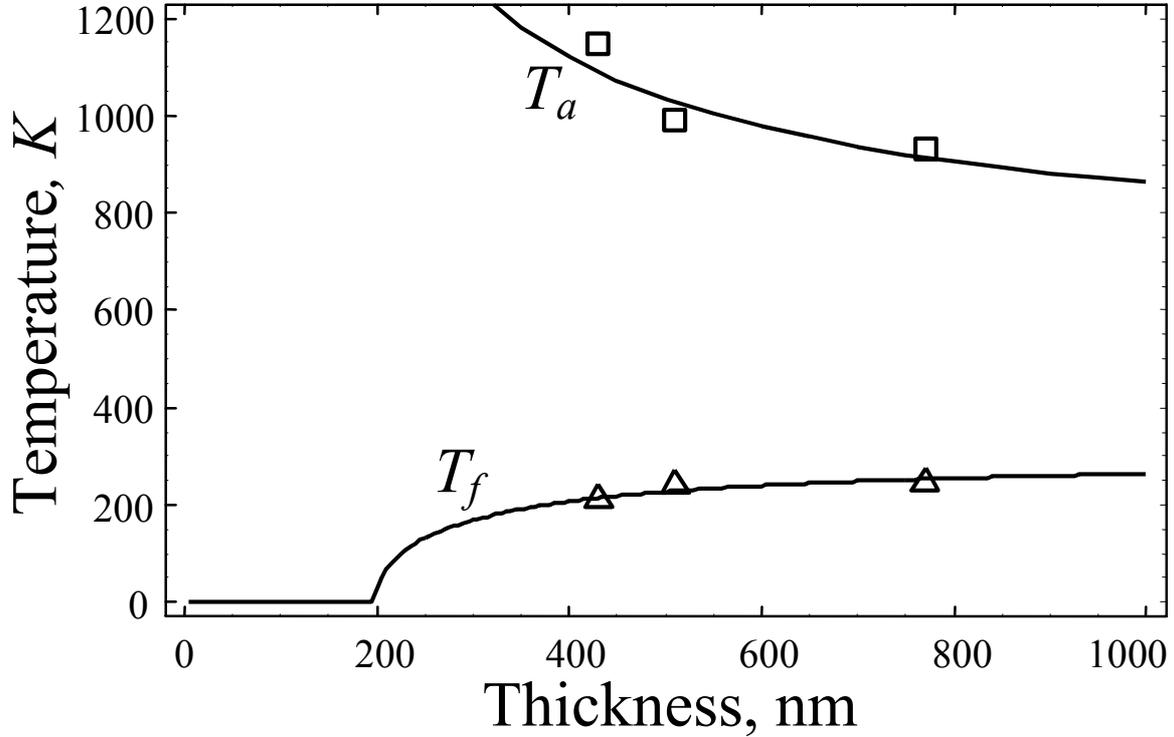

**Figure 6**. Comparison of the thickness dependence of the freezing temperature (triangles) and activation energy (boxes) obtained from the analysis of the experimental data [6] with theoretical dependences (solid lines) for the following parameters $E_a/k_B = 690\ K$, $T_d = 620\ K$.

Several results observed in [6] are in a good agreement with our theoretical predictions. Namely, piezoelectric activity of PMN thin films prior of application of external electric field is the manifestation of existence of built-in internal field related to the mismatch effect, which is the true reason of self polarization effect discussed by the authors of paper [6]. Really experimentally observed the external electric field threshold value which is necessary to induce the ferroelectric long range order for PMN is about 1.5 kV/cm [29]. Our fitting to the experimental data [6] gives the following expression for the internal field $d^* E_m(h) = (238 \cdot 10^{-23} J \cdot \mu m)/h$ with thickness $h$ in micrometers (since length $l_d$ is of several lattice constant order, we neglected the term $2l_d^2$ in comparison with $h(\lambda + l_d)$). Taking into account that the displacement of ferroelectric active ions in lead containing relaxors is of order $10^{-10} \div 10^{-11}\ m$ (see e.g. [31], [32]) the dipole moment can be evaluated as $d \cong 5(10^{-29} \div 10^{-30}) C\, m$. Then the effective dipole moment $d^*$ one can evaluate as atomic value $d$ multiplied by the factor of 100 [17], [25]. Therefore for the films investigated in [6] we obtained $E_m(h \sim 500\,nm) \cong (100 \div 10)\,kV/cm$ which is much higher than threshold value.

The main differences between films and bulk of relaxor ferroelectrics, e.g. lower permittivity, shift of its maximum to the lower temperatures with film thickness decrease, clearly follow from comparison of corresponding curves in Figs. 3c and 3d as well as in Figs. 4c and 4d. One can also see that for the strongly strained films the abovementioned differences between permittivity of film and bulk become more pronounced. On the other hand one can expect the essential increase of dielectric permittivity for the very thin films on the substrate with smallest possible misfit between them. It is because in this case the film is very close to the free standing one, for which theory forecasts the strong increase of permittivity in the vicinity of size driven phase transition (see Fig. 2). The compensation of the built-in internal field by the external one will lead to the enhancement of dielectric permittivity as well as to the decrease of order parameter.

Therefore the proposed theory of relaxor ferroelectric films explains main experimental data for PMN films. The further experimental investigations of size effects of thin films of relaxor ferroelectrics in the wide range of thickness are extremely desirable.

## REFERENCES

[1] S.W. Choi et al., Ferroelectrics 100, 29 (1990).

[2] P.E. Park and T.R. Shrout, J.Appl.Phys. **82**, 1804 (1997).

[3] M.Tyunina, J.Levoska, A. Sternberg, S.Leppavuori, J. Appl. Phys. **84**, 6800 (1998).

[4] V.Bornard, S.Trolier-McKinstry, J. Appl. Phys. **87**, 3958 (2000).

[5] V.Bornard, S.Trolier-McKinstry, K.Takemura, C.A.Randal, J. Appl. Phys. **87**, 3965 (2000).

[6] Z.Kighelman, D. Damjanovic and N. Setter, J. Appl. Phys. **89**, 1393 (2001).

[7] M.Tyunina, J.Levoska, Phys. Rev. B **63**, 224102 (2001).

[8] M.Tyunina, J.Levoska, S.Leppavuori, J. Appl. Phys. **91**, 9277 (2002).

[9] M.Tyunina, J.Levoska, K.Kundzinsh and V. Zauls, Phys. Rev. B **69**, 224101 (2004).

[10] M.Tyunina, J.Levoska, Phys. Rev. B **70**, 132105 (2004).

[11] S. Kamba, M. Kempa, V. Bovtun, J.Petzelt, K.Brinkman and N. Setter, J. Phys.: Condens. Matter **17**, 3965 (2005).

[12] M.D.Glinchuk, E.A.Eliseev, V.A.Stephanovich, Physica B, **332**, 356 (2002).

[13] M.D.Glinchuk, E.A.Eliseev, V.A.Stephanovich, R.Farhi, J. Appl. Phys. **93**, 1150 (2003).

[14] M.D.Glinchuk, A.N.Morozovska, J. Phys.: Condens. Matter **16**, 3517 (2004).

[15] Y.G. Wang, W.L. Zhong and P.L. Zhang, Phys. Rev. B **51**, 5311 (1995).

[16] M.D.Glinchuk, B.Hilcher, V.A.Stephanovich, K.Weron, Phase Transitions **76**, 557 (2003).

[17] M.D.Glinchuk, R.Farhi, J. Phys.: Cond. Matter. **8**, 6985 (1996).

[18] M.D.Glinchuk, E.A.Eliseev, V.A.Stephanovich, Ferroelectrics **298**, 69 (2004).

[19] M.D.Glinchuk, A.N.Morozovska, E.A.Eliseev, Integrated Ferroelectrics **64**, 17 (2005).

[20] M.D.Glinchuk, A.N.Morozovska, E.A.Eliseev, E-print: http://arxiv.org/cond-mat/0504537 (2005).

[21] N.A.Pertsev, A.G.Zembilgotov, A.K.Tagantsev, Phys. Rev. Lett. **80**, 1988 (1998).

[22] J.S.Speck, W.Pompe, J.Appl.Phys. **76**(1), 466 (1994).

[23] E.A.Eliseev, M.D.Glinchuk, Physica Status Solidi (b) **241**(11), R52 (2004)

[24] E.A.Eliseev, M.D.Glinchuk, Physica Status Solidi (b) **241**(15), 3495 (2004).

[25] M.D.Glinchuk, V.A.Stephanovich, J. Phys.: Cond. Matter. **6**, 6317 (1994).

[26] E.A.Eliseev, M.D.Glinchuk, E-print: http://arxiv.org/cond-mat/0412397 (2004).

[27] L.Landau, E.Lifshits, *Electrodynamics of continuous media*, Pergamon Press, Oxford (1984).

[28] M.D.Glinchuk, British Ceramic Transactions. **103**, 76 (2004).

[29] E.V.Colla, N.K. Yushin and D. Viehland, J. Appl. Phys. **83**, 3298 (1998).

[30] M.D.Glinchuk, V.A.Stephanovich, J. Appl. Phys. **85**, 1722 (1999).

[31] V.V. Laguta, M.D. Glinchuk, S.N. Nokhrin, I.P. Bykov, R. Blinc A. Gregorovic and B. Zalar, Phys. Rev. B **67**, 104106 (2003).

[32] V.V. Laguta, M.D. Glinchuk, I.P. Bykov, R. Blinc and B. Zalar, Phys. Rev. B **69**, 054103 (2004).